\documentclass[aps,prl,twocolumn,showpacs,groupedaddress]{revtex4-1}
\usepackage{pifont}
\usepackage{amsmath}
\usepackage{graphicx}
\usepackage{amssymb}
\usepackage{amsfonts}
\usepackage{subfigure}
\usepackage{booktabs}
\usepackage{setspace}
\usepackage{threeparttable}
\usepackage{enumerate}
\usepackage{wasysym}

\begin{document}

\title{Itinerant chiral ferromagnetism in a trapped Rashba spin-orbit coupled Fermi gas}
\author{Shang-Shun Zhang$^{1,2}$}
\author{Wu-Ming Liu$^{1}$}
\author{Han Pu$^{2,3}$}
\affiliation{ $^{1}$Beijing National Laboratory for Condensed Matter Physics, Institute of
Physics, Chinese Academy of Sciences, Beijing 100190, China \\
$^{2}$ Department of Physics and Astronomy, and Rice Center for Quantum Materials, Rice University, Houston, Texas 77251, USA \\
$^3$
Center for Cold Atom Physics, Chinese Academy of Sciences, Wuhan 430071, China}
\date{\today}

\begin{abstract}
How ferromagnetic phases emerge in itinerant systems is an outstanding problem in quantum magnetism. Here we consider a repulsive two-component Fermi gas confined in a two dimensional isotropic harmonic potential and subject to a large Rashba spin-orbit (SO) coupling, whose single-particle dispersion can be tailored by adjusting the SO coupling strength. We show that the interplay among SO coupling, correlation effects and mean-field repulsion leads to a competition between ferromagnetic and non-magnetic phases. At intermediate interaction strengths, ferromagnetic phase emerges which can be well described by the mean-field Hartree-Fock theory; whereas at strong interaction strengths, a strongly correlated non-magnetic phase is favored due to the beyond-mean-field quantum correlation effects. Furthermore, the ferromagnetic phase of this system possesses a chiral current density induced by the Rashba spin-orbit coupling, whose experimental signature is investigated.
\end{abstract}


\maketitle
\textit{Introduction ---} Itinerant ferromagnetism represents an outstanding problem in many-body physics. Based on Stoner's argument, a fermionic system in continuum may become ferromagnetic when the repulsive interaction strength exceeds a critical value \cite{stoner}. In the context of a spin-1/2 Fermi gas, ferromagnetism means that the two spin species tend to phase separate to form spin domains as such a configuration obviously reduces interaction energy. Attempt to realize ferromagnetic state in repulsive Fermi gas was made by the MIT group in 2009 \cite{ferro}. Although some indirect evidences were present, spin domain formation was not observed. Later it was clarified that their system suffers from strong atom loss as the atoms tend to form tightly bound dimers, and ferromagnetism was therefore not expected \cite{noferro}. From perhaps a more fundamental point of view, even if a repulsive Fermi gas is stable, it is not completely clear whether a ferromagnetic state will result. This is because the Stoner's criterion is based on a mean-field argument, in which ferromagnetism arises once the mean-field repulsion overcomes the kinetic energy. However, it has been conjectured that, under strong repulsive interaction, the Fermi gas may form a strongly correlated non-magnetic state \cite{gutz,zhai} that competes with the ferromagnetic state. Here the quantum correlation effects, neglected in the mean-field argument, play a more dominant role. Therefore whether itinerant ferromagnetic phases can exist in repulsive Fermi gases remains as an open question.

In this Letter, we show that itinerant ferromagnetism {\em can exist} in a repulsive Fermi gas subject to spin-orbit (SO) coupling \cite{SOC1,SOC11,SOC2,SOC3,SOC4,SOC5,SOC6}. Itinerant ferromagnetism is a consequence of the interplay among kinetic energy, mean-field repulsion between the spin species, and quantum correlation effects. The key here is that the SO coupling significantly modifies the single-particle dispersion of the Fermi gas such that a ferromagnetic state emerges without the need of a very strong repulsive interaction.   

More specifically, we consider here a Rashba SO coupled repulsive two-component Fermi gas confined in a two dimensional (2D) isotropic harmonic potential. The trapping potential is necessary for any cold atom experiment as it provides atomic confinement. However, in the current situation, it plays an additional role: Together with the Rashba SO coupling, it produces a Landau level-like single-particle spectrum whose band flatness can be controlled by the SO coupling strength \cite{wu,HanPu1,HanPu2}, which as we will show is crucial for the existence of the itinerant ferromagnetic phases in our system. If the system consists of a single spin-1/2 particle, the physics is well understood. The ground state is represented by a half vortex state \cite{wu}. An interesting feature of the system is that, under the limit of large SO coupling strength, the single-particle spectrum exhibits Landau level-like structure. In the case of an ensemble of spin-1/2 bosons, as previous works have shown \cite{HanPu1,HanPu2}, the near flat band structure leads to intriguing spin textures and strongly correlated phases. Our current work tries to answer the question: What happens when we have an ensemble of repulsive spin-1/2 fermions? 


\begin{figure}[!h]
\begin{center}
\includegraphics[width=2.5in]{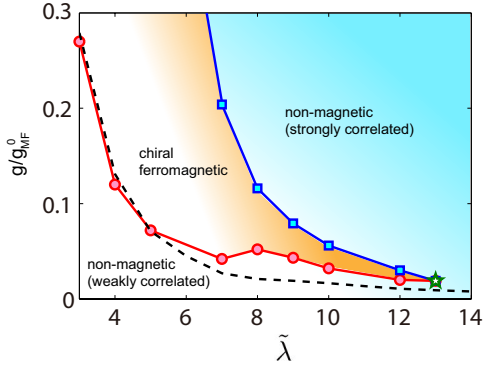}
\end{center}
\caption{(Color online) Phase diagram in the $g$-$\tilde{\lambda}$ plane. Here we consider 6 repulsively interacting spin-1/2 fermions confined in a 2D harmoinc trap, subject to Rashba SO coupling. $g$ is the interaction strength, and is normalized to $g_{\rm MF}^0=2\pi\hbar^2/M$ which is the mean-field critical interaction strength for a 2D Fermi gas without SO coupling. $\tilde{\lambda}$ is the dimensionless SO coupling strength. We can see that the phase diagram contains three phases: the weakly correlated non-magnetic phase, chiral magnetic phase and strongly correlated non-magnetic phase. The dashed line represents the mean-field results which contains only two regimes: a non-magnetic phase below the dashed line and a ferromagnetic phase above the dashed line.}
\label{fig:phase}
\end{figure}

To address this problem and to elucidate the relationship between ferromagnetism and interaction effects, we carried out a fully quantum mechnical exact diagonalization (ED) calculation combined with a mean-field Hartree-Fock calculation. These two complementary methods allow us to roughly divide the interaction effects into two parts: (1) it leads to a mean-field repulsion between the two spin species; (2) it builds up quantum correlations in the system. We will show that the former favors ferromagnetism, whereas the latter has an opposite effect. The competition between them gives rise to the phase diagram shown in Fig.~\ref{fig:phase}, where a ferromagnetic phase occupies a finite region in the parameter space spanned by the interaction strength and the SO coupling strength.

\textit{The model ---}
We consider a spin-1/2 Fermi gas, with atomic mass $M$ and chemical potential $\mu$, confined in the $x$-$y$ plane by an isotropic harmonic trap $V(r)={1\over 2}M\omega r^2$ ($r=\sqrt{x^2+y^2}$), subject to a Rashba SO coupling $\mathcal{V}_{\rm soc}=\lambda (p_y\sigma_x-p_x\sigma_y)$, where $\sigma_{x,y}$ are Pauli matrices. The model Hamiltonian is given by $\mathcal{H}=\mathcal{H}_0 + \mathcal{H}_{\rm int}$ where
\begin{eqnarray}
\mathcal{H}_0 = \int d^{2} \vec{r}\, \Psi^{\dagger} \left[ \frac{ -\hbar^{2} \nabla^{2}}{ 2M}-\mu + \mathcal{V}_{\rm soc}+V(r) \right] \Psi,
\end{eqnarray}
with $\Psi = (\Psi_\uparrow, \Psi_\downarrow)^T$ being the atomic field operator, is the single-particle Hamiltonian, and
\begin{eqnarray}
\mathcal{H}_{\rm int} = g \int d^{2} \vec{r}\,  \Psi^{\dagger}_{\uparrow}( \vec{r} )\Psi^{\dagger}_{\downarrow}(  \vec{r} )
     \Psi_{\downarrow}(  \vec{r} )\Psi_{\uparrow}(\vec{r} ),
\end{eqnarray}
with $g>0$ decribes repulsive $s$-wave contact interaction. In what follows, we will adopt the trap units where the units for length and energy are given by $a_{\rm ho}=\sqrt{\hbar/(M\omega)}$ and $\hbar \omega$, respectively. Under this unit system, the interaction strength $g$ has be units of $\hbar \omega a_{\rm ho}^2 = \hbar^2/M$. We also define a dimensionless SO coupling strength $\tilde{\lambda} = M\lambda a_{\rm ho}/\hbar^2$. In the limit $\tilde{\lambda} \gg 1$, the single-particle spectrum exhibits Landau level-like structure and the curvature of each Landau band is proportional to $1/\tilde{\lambda}^2$, which provides a way to control the band flatness. Flat band structure will have two effects on an interacting many-body system: On the one hand, it may reduce the critical interaction strength for the ferromagnetic transition according to the Stoner's criterion. On the other hand, it makes quantum correlation more pronounced. Which of these two effects become more dominant determines whether the system is ferromagnetic or not.

\begin{figure}[!t]
\begin{center}
\includegraphics[width=2.8in]{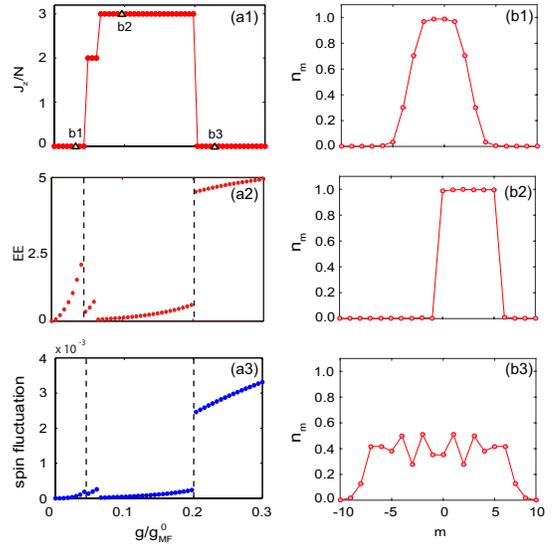}
\end{center}
\caption{(Color online) Left column: (a1)-(a3) show the ground state angular momentum per particle $J_z/N$, entanglement entropy (EE) and ground state spin fluctuations $(\Delta S_z)^2$ as a function of interaction strength for $\tilde{\lambda}=7, N=6$. The non-zero $J_z/N$ indicates a magnetic ground state. Right column: (b1)-(b3) show three representative single-particle occupation number $n_m$ of state $|m\rangle$ for interaction strengthes marked by the yellow triangles in (a1).}
\label{fig:groundstate}
\end{figure}

{\em ED results ---}
The Landau level structure of the single-particle spectrum allows us to use the ED method to study a few-body system, where we restrict our calculation to the lowest Landau level (LLL). The single-particle Hamiltonian $\mathcal{H}_0$ conserves the total angular momentum $J_z$, which is the sum of the orbital and the spin angular momentum. Single-particle states in the LLL can be labeled by a single quantum number $|m \rangle$, whose total angular momentum is $J_z=m+1/2$, and whose energy (apart from a constant) is approximately $m(m+1)/\tilde{\lambda}^2$ \cite{wu,HanPu1,HanPu2}. A set of such states form a Fock space basis, upon which the total Hamiltonian can be expanded~\cite{HanPu2}. For details, see Supplemental Material~\cite{supp}.

We present our ED results for a system of $N=6$ fermions. Under the total Hamiltonian $\mathcal{H}$, $J_z$ of the whole system remains as a good quantum number. Figure~\ref{fig:groundstate}(a) displays $J_z$ of the ground state as a function of the interaction strength $g$ for a fixed SO coupling strength $\tilde{\lambda}=7$. As one can see, $J_z=0$ for small $g$, becomes finite for intermediate $g$, and vanishes again at large $g$. The single-particle occupation number $n_m=\langle a_m^\dag a_m \rangle$, with $a_m$ the annihilation operator associated with state $|m \rangle$, for the three representative cases are plotted in Fig.~\ref{fig:groundstate}(b1)-(b3) as a function of $m$. For a weakly interacting system, as shown in Fig.~\ref{fig:groundstate}(b1), interaction induces a few particle-hole excitations near the ``Fermi surface''. However, the ground state still preserves the time reversal symmetry, i.e., $n_{m}=n_{-m-1}$. The density profiles for the two spin species are identical. One can also calculate the local spin vector $\vec{s}(\vec{r}) = \langle \Psi^\dag(\vec{r}) \vec{\sigma} \Psi(\vec{r}) \rangle$ and show that it vanishes everywhere. Hence the state is  a non-magnetic state.

\begin{figure}[!t]
\begin{center}
\includegraphics[width=3.2in]{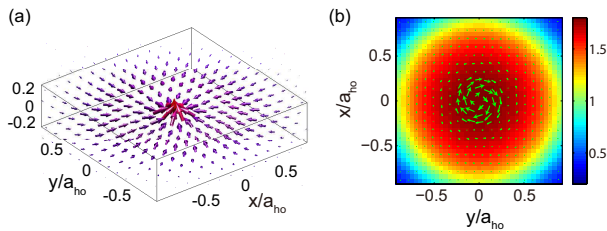}
\end{center}
\caption{(Color online) (a) shows the spin texture of the ferromagnetic state and (b) shows the total number density (background color) and the chiral current density (arrows) of the ferromagnetic state.}
\label{fig:spincurrent}
\end{figure}

At intermediate interaction strength, as shown in Fig.~\ref{fig:groundstate}(b2), the ground state breaks time reversal symmetry with $n_{m} \neq n_{-m-1}$, the two spin species possess non-overlapping density profiles, and non-vanishing local spin vector $\vec{s}(\vec{r})$ emerges, see Fig.~\ref{fig:spincurrent}(a). This indicates that the state is a ferromagnetic state. Furthermore, we calculated the current density of this state. With Rashba SO coupling, the current density is given by
\begin{eqnarray}
\vec{j}(\vec{r})=\sum_m \vec{j}_{\rm orbit}^m n_m+\tilde{\lambda} \hat{z}\times \vec{s}\,,
\label{current}
\end{eqnarray}
where $\vec{j}_{\rm orbit}^m={i}[(\nabla \phi_m^{\dagger})\phi_m-\phi_m^{\dagger}\nabla \phi_m]/2$ comes from the orbital motion where $\phi_m$ represents the wave function of the single-particle state $|m \rangle$.
Due to the the time reversal symmetry of the Hamiltonian, we have $\vec{j}_{\rm orbit}^m=-\vec{j}_{\rm orbit}^{-m-1}$.
In the non-magnetic state, $n_m=n_{-m-1}$ and $\vec{s}=0$, both terms on the right hand side of Eq.~(\ref{current}) vanish. However, for the ferromagnetic state, they are both finite and leads to a chiral current as shown in Fig.~\ref{fig:spincurrent}(b). As a result, we call the magnetic state chiral ferromagnetic.

At large interaction strength, as shown in Fig.~\ref{fig:groundstate}(b3), the time reversal symmetry is restored, and once again we have $\vec{s}(\vec{r}) = 0$ and $\vec{j}(\vec{r}) = 0$ as in the weakly interacting regime. The fluctuations of $n_m$ indicates that this non-magnetic state is strongly correlated. To quantify the quantum correlation and fluctuation, we calculated the entanglement entropy (EE) of the system \cite{supp}, and the total spin fluctuation $(\Delta S_z)^2= \langle \hat{S}_z^2 \rangle - \langle \hat{S}_z \rangle^2$, and plot them as  functions of $g$ in Fig.~\ref{fig:groundstate}(a2) and (a3), respectively. Both EE and $(\Delta S_z)^2$ for the large interaction regime are significantly higher than those in the other two regimes.

With the above results and similar calculations for other SO coupling strengths, we can present the phase diagram as shown in Fig.~\ref{fig:phase}. For $\tilde{\lambda} \apprle 13$, there exists a window of ferromagnetic phase at intermediate values of $g$. As $g$ increases from zero to a lower critical value (represented by the red solid line with filled circles), the weakly correlated non-magnetic state becomes ferromagnetic. Note that this lower critical value is much smaller than $g^0_{\rm MF}=2\pi\hbar^2/M$, the mean-field ferromagnetic critical interaction strength of a 2D homogeneous Fermi gas without SO coupling \cite{conduit,toigo}. This can be understood from the Stoner's argument and the flat band single-particle spectrum. In Ref.~\cite{pilati}, it was shown that the critical interaction strength for ferromagnetic transition in a repulsive Fermi gas can also be reduced by adding a weak optical lattice, as the lattice potential helps to quench the kinetic energy. The essential physics here is similar to our situation. Howver, as $g$ further increases to an upper critical value (represented by the blue solid line with empty squares), the ferromagnetic state gives its way to a strongly correlated non-magnetic state. As $\tilde{\lambda}$ increases, i.e., the single-particle band becomes flatter, this window of ferromagnetic phase shrinks quickly, and eventually vanishes for $\tilde{\lambda} \apprge 13$. At such large SO coupling strength, the single-particle band becomes so flat that a very small interaction strength gives rise to strong correlations that disfavor the ferromagnetic state.

To examine the finite-size effect, we made ED calculations for $N=4,6,8$, and found that with increasing $N$, the quantum correlation effect is somewhat weakened. For fixed $\tilde{\lambda}$, the lower critical interaction strength at which the weakly correlated non-magnetic phase changes to the ferromagnetic phase is not very sensitive to $N$, while the upper critical interaction strength at which the ferromagnetic phase becomes the strongly correlated non-magnetic phase increases with $N$. Furthermore, the critical SO coupling stength at which the ferromagnetic window vanishes also increases with $N$. As a result, the ferromagnetic regime in the $g$-$\tilde{\lambda}$ phase diagram is enlarged with increasing $N$.


\textit{Hartree-Fock results ---} If it is the correlation effects that destroy the ferromagnetic state, then one should not expect this to occur in a mean-field theory, which neglects quantum correlation. To examine this, we now turn to a mean-field Hartree-Fock (HF) calculation.
Under the HF theory, the many-body wave function takes the form: $\Psi_{\rm HF}={1\over \sqrt{N!}} \sum_P (-1)^{P} \phi_1(\vec{r}_1)\phi_2(\vec{r}_2)...\phi_N(\vec{r}_N)$, 
where $P$ represents permutations, and $\phi_\alpha$'s are single-particle orbitals that satisfy the following HF equations:
\begin{eqnarray}
&&\left[  -\frac{1}{2}\nabla ^{2}+i\tilde{\lambda}\left( -\partial _{y}\sigma
_{x}+\partial _{x}\sigma _{y}\right) +\frac{1}{2}r^{2} \right. \nonumber \\
&&\;\; + \left. \frac{g }{4}%
n\!\left( r\right) -\frac{g }{4}\vec{m}\!\left( r\right) \cdot \vec{\sigma}%
\right] \phi _{\alpha }\!\left( \vec{r}\right) =\xi _{\alpha }\,\phi _{\alpha }\!\left(
\vec{r} \right)\,,
\end{eqnarray}
where $ n\!\left( r\right) =\sum_{\alpha =1}^{N}|\phi _{\alpha }|^2$ and $ \vec{m}\left( r\right) = \sum_{\alpha =1}^{N}\phi _{\alpha }^{\dag } \vec{
\sigma}\,\phi _{\alpha }$
are local density and spin vector, respectively.

We numerically solve the HF equations self-consistently {\em without} invoking the LLL approximation (for details, see Supplemental Material~\cite{supp}). In Fig.~\ref{fig:hartree} we plot density profiles from this calculation. Here we also take $N=6$ and $\tilde{\lambda}=7$ in order to make comparisons with the ED results. However, we also performed HF calculations up to $N=200$ and found no qualitative differences from the $N=6$ results presented here. For small interaction strength $g=0.018 g^0_{\rm MF}$ [Fig.~\ref{fig:hartree}(a)], both HF and ED tell us that the state is non-magnetic with identical density profiles for both spin species. Furthermore, the results from the two theories agree with each other very well.  At $g=0.03 g^0_{\rm MF}$ [Fig.~\ref{fig:hartree}(b)], ED predicts a non-magnetic state, whereas HF indicates that the system already enters the ferromagnetic regime. In fact, HF calculation predicts a critical interaction strength $g_{\rm HF} \approx 0.027g^0_{\rm MF}$, while the corresponding critical interaction strength for ED is $g_{\rm ED} \approx 0.05 g^0_{\rm MF}$. That $g_{\rm ED} > g_{\rm MF}$ can be attributed to the fact that the quantum correlation in the ED calculation disfavors the ferromagnetic phase. At $g= 0.173 g^0_{\rm MF}$ [Fig.~\ref{fig:hartree}(c)], HF and ED agree with each other again, both predicting a ferromagnetic state. At a large interaction strength $g=0.234 g^0_{\rm MF}$ [Fig.~\ref{fig:hartree}(d)], discrepancies arise between the two calculations again: ED predicts a non-magnetic state, while HF gives a ferromagnetic state. In fact, as we have expected, for $g> g_{\rm HF}$, HF always predicts a ferromagnetic state. In contrast, our ED calculation shows that for sufficiently large $g$, strong correlation destroys the ferromagnetic state. For the parameters used in Fig.~\ref{fig:hartree}, ED shows that ferromagnetic phase only exists for $0.05 \apprle g/g_{\rm MF}^0 \apprle 0.2$. 

\begin{figure}[!h]
\begin{center}
\includegraphics[width=3in]{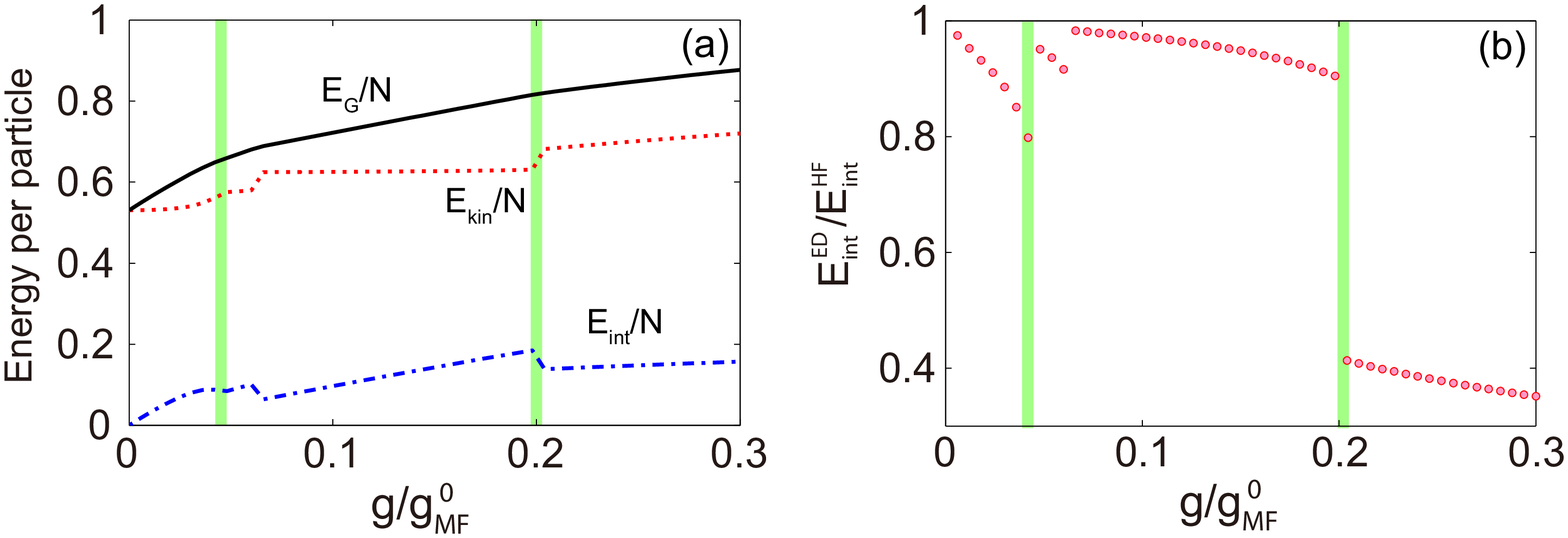}
\end{center}
\caption{(Color online) (a) ED results for the ground state energy $E_{G}$, the kinetic energy $E_{\rm kin}$, and the interaction energy $E_{\rm int}$ as functions of interaction strength. (b) The ratio of the interaction energy from the ED calculation and that from the HF calculation. The two vertical lines separate the parameter space into three phases according to the ED calculation: from left to right, we have the weakly correlated non-magnetic phase, the ferromagnetic phase, and the strongly correlated non-magnetic phase. Here $N=6$ and $\tilde{\lambda}=7$. }
\label{fig:int}
\end{figure}

In the phase diagram of Fig.~\ref{fig:phase}, the dashed line represents $g_{\rm HF}$, which separates the phase space into non-magnetic (below the dashed line) and ferromagnetic regimes (above the dashed line). $g_{\rm HF}$ decreases quickly as $\tilde{\lambda}$ increases (which can again be understood as due to the band flattening), but never terminates as in the case of ED. To further demonstrate the effects of quantum correlation, we plot in Fig.~\ref{fig:int} the energy as a function of interaction strength at $\tilde{\lambda}=7$. Figure \ref{fig:int}(a) shows how the total energy $E_G$, the kinetic energy $E_{\rm kin}$ and the interaction energy $E_{\rm int}$ from the ED calculation change as $g$. As $g$ increases, $E_G$ keeps increasing monotonically, while $E_{\rm int}$ decreases at both phase transition points (shown by the vertical lines) at the cost of increasing $E_{\rm kin}$. In Fig.~\ref{fig:int}(b) we plot the ratio of the interaction energy, which is simply $g$ times the density-density correlation between the two spin species integrated over all space, from the ED and the HF calculation. As it shows, in the weakly correlated non-magnetic and the ferromagnetic regimes, the ED and the HF results are comparable to each other. By contrast, in the strongly correlated regime, the interaction energy from the ED calculation is signaficantly lower than that from the HF calculation. This clearly shows how the system develops nontrivial quantum correlations such that, even though the density profiles of the two spin species completely overlap with each other, the joint probability of finding two unlike spins at the same position is strongly suppressed.


\begin{figure}
\begin{center}
\includegraphics[width=2.7in]{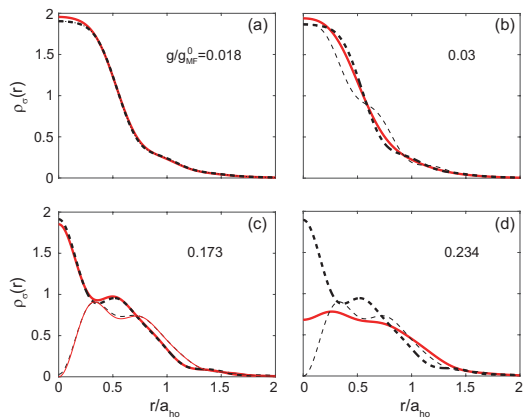}
\end{center}
\caption{(Color online) Density profiles of each spin species for different interaction
strengths with $N=6$ and $\tilde{\lambda}=7$, by both ED (red solid lines: thick lines for spin up and thin lines for spin down) and HF (black dashed lines: thick lines for spin up and thin lines for spin down) methods. }
\label{fig:hartree}
\end{figure}

Finally, we propose an experimental procedure to detect the chrial current associated with the ferromagnetic state. The procedure goes as follows: First the ground state (either magnetic or not) is prepared. Then the harmonic trap is suddently distorted from isotropic to anisotropic. For a non-magnetic state, as shown in the upper panel of Fig.~\ref{fig:dynamic} obtained from a time-dependent HF calculation \cite{supp}, this induces a quadrupole mode. By contrast, for an initial chiral ferromagnetic state, the whole cloud also undergoes an angular rotation, analogous to the scissors mode in a condensate with vortices \cite{sci1}.

\textit{Conclusion ---}
In summary, we have shown that how the Landau level-like band structure of a 2D Rashba SO coupled Fermi gas, with a controllable band flatness, can be exploited to exhibit itinerant ferromagnetism. The near-flat band structure dramatically reduces the critical interaction strength required for the ferromagentic phase transition. We employed two complementary methods, the fully quantum ED method and the mean-field HF method, to investigate this problem. Our calculation elucidates the interplay between the mean-field repulsion and the quantum correlation effects, and shows that the former favors while the latter tends to destroy ferromagnetism. The emergence and disappearance of the ferromagnetic phase result from the competition between these two factors. We have also shown that the ferromagnetic phase in our system is accompanied by a chiral density current resulting from the SO coupling. This chiral density current and the spin texture that characterize the itinerant ferromagnetic state can be readily detected using today's cold atom techniques. We hope our work may open new avenues of research in both SO coupling and itinerant magnetism in cold atoms.

Finally, we comment that Dresselhaus SO coupling has recently been realized by the Shanxi group \cite{shanxi}. The single-particle spectrum of a harmocally trapped 2D spin-1/2 particle remains exactly the same if the Rashba SO coupling is changed to the Dresselhaus coupling. Our results for the repulsive Fermi gas remain essentially the same under Dresselhaus coupling \cite{supp}. 

\begin{figure}[!t]
\begin{center}
\includegraphics[width=2.7in]{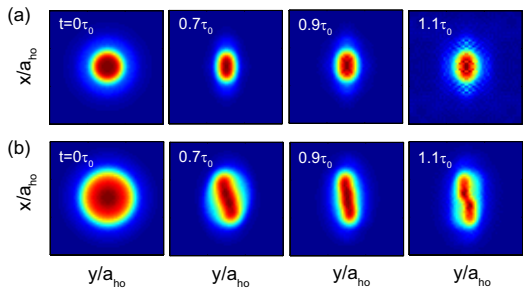}
\end{center}
\caption{(Color online) Time evolution of the atomic cloud after a suddent trap deformation. At $t=0$, the trapping frequency along the $y$-axis is suddently changed from $\omega$ to $3.16\omega$, while that along the $x$-axis remains at $\omega$. The upper (lower) panel shows the dynamics of a non-magnetic (ferromagnetic) state. Here $N=6$, $\tilde{\lambda}=7$ and $\tau_0=1/\omega$.}
\label{fig:dynamic}
\end{figure}

\begin{acknowledgments}
HP acknowledges support from the US NSF and the Welch Foundation (Grant No. C-1669), and WML is supported by the NKBRSFC under grants Nos. 2011CB921502, 2012CB821305, NSFC under grants Nos. 61227902, 61378017, 11434015, and SPRPCAS under grants No. XDB01020300.
\end{acknowledgments}

\appendix

\begin{widetext}
\section{Supplementary Materials for ``Itinerant chiral ferromagnetism in a trapped Rashba spin-orbit coupled Fermi gas''}

In this Supplementary Materials, we provide more technical details of the main manuscript.

\subsection{Exact diagonalization scheme}
For a single spin-1/2 particle with Rashba spin-orbit (SO) coupling confined in a two dimensional (2D) isotropic harmonic trap, the single-poarticle Hamiltonian is given by $\mathcal{H}_0$ in the main text. The spatial wave function of the eigenstates take the form:
\begin{equation}
\label{single}
\Phi_{n,m}(\vec{r}) = \left( \begin{array}{c}f_{nm}(r) \\ g_{nm}(r) e^{i\phi} \end{array} \right)\,e^{im\phi} \,,\;\;\Phi_{n,-m-1}(\vec{r}) = \left( \begin{array}{c} g_{nm}(r) e^{-i\phi} \\-f_{nm}(r)  \end{array} \right)\,e^{-im\phi}\,,\;\;\;\;n=0, 1, 2, ...; \, m=0, \pm 1, \pm 2, ...
\end{equation}
which form a degenerate time reversed pair with eigenenergies $\epsilon_{n,m}=\epsilon_{n,-m-1}$. In the limit that the dimensionless SO coupling strength $\tilde{\lambda} \gg 1$, the eigenenergies (apart from a constant) take the following approximate form:
\[ \epsilon_{n,m}=\epsilon_{n,-m-1} = \left[ n + \frac{m(m+1)}{\tilde{\lambda}^2} \right] \,\hbar \omega \,.\]
For a few-body system with weak interaction and small particle number, the Hilbert space is limited to the lowest Landau level (LLL) which is specified by the quantum number $n=0$. We introduce a cutoff $m^*$ which further reduces the Hilbert space to that with $-m^*-1<m<m^*$. The value of $m^*$ is determined by specific values of $N$ and $g$. Given $N$ fermionic particles filled to $M=2m^*+2$ single particle states, we obtain totally ${M!\over N! (M-N)!}$ Fock states. Due to the rotational symmetry of this system, we are able to divide the full truncated Hilbert space into several independent subspaces with fixed total angular momentum $J_z=\sum_{i=1}^N (m_i+{1\over 2})$, which considerably reduces the dimension of the Hamiltonian that needs to be diagonalized.

Next we present the main steps for the ED scheme for specific subspace with $J_z$ and particle number $N$. The Fock states are denoted by $|p_i\rangle=a_{m_1}^{\dagger} a_{m_2}^{\dagger} ... a_{m_N}^{\dagger}|0\rangle, i=1,2, ..., D$ with the convention $m_1<m_2<...<m_N$.
For later use, we associate each occupied single particle-state $m_i$ with a number $N_{m_i}$ (for example, $N_{m_1}=1$).
The single-particle part of the Hamiltonian $\mathcal{H}_0$ is diagonal under this basis:
\begin{eqnarray}
\langle p_i | \mathcal{H}_0 | p_j \rangle =\sum_{\alpha=1}^N \epsilon_{m_{\alpha}} \delta_{ij}\,.
\end{eqnarray}
where $\epsilon_m = \epsilon_{n=0,m}$.
Under the same basis, the diagonal matrix elements of the interacting Hamiltonian $\mathcal{H}_{\rm int}$ take the form:
\begin{eqnarray}
\langle p_i | \mathcal{H}_{\rm int} | p_i \rangle ={g\over 4} \int d^2\vec{r} \,(\rho^2-\vec{s}^2)\,,
\end{eqnarray}
where $\rho(r)=\langle p_i| \hat{\rho}|p_i\rangle$ and $\vec{s}=\langle p_i| \hat{\vec{s}}|p_i\rangle$ represent the local density and spin vector, respectively. This diagonal matrix elements can be regarded as the mean-field Hartree-Fock interaction energy associated with the Fock state $|p_i \rangle$.  The non-diagonal matrix elements of $\mathcal{H}_{\rm int}$ are non-vanishing only between two Fock states that differ by two single-particle states, say $|p\rangle=...a_m^{\dagger}...a_n^{\dagger}...|0\rangle$ and $|q\rangle=...a_k^{\dagger}...a_l^{\dagger}...|0\rangle$ with the constraint $m+n=k+l$:
\begin{eqnarray*}
\langle q | \mathcal{H}_{\rm int} | p \rangle =(-1)^{N_m+N_n+N_k+N_l}g\int d^2\vec{r} \left[\Psi_{l\uparrow}^*\Psi_{k\downarrow}\Psi_{m\downarrow}\Psi_{n\uparrow}+ \Psi_{k\uparrow}^*\Psi_{l\downarrow}\Psi_{n\downarrow}\Psi_{m\uparrow}
-\Psi_{k\uparrow}^*\Psi_{l\downarrow}\Psi_{m\downarrow}\Psi_{n\uparrow}
-\Psi_{l\uparrow}^*\Psi_{k\downarrow}\Psi_{n\downarrow}\Psi_{m\uparrow}\right]\,,
\end{eqnarray*}
where $\Psi_{m\sigma},\sigma=\uparrow,\downarrow$ denotes the wave function of the single particle state in the lowest Landau level. The non-diagonal part of $\mathcal{H}_{\rm int}$ builds up correlations between different Fock states. It mixes Fock states with different spin polarization, therefore tends to suppress the magnetic phase.

\subsection{Calculation of entanglement entropy (EE)}
Entanglement measure is useful to analyze correlation properties of the ground state. We calculate EE in the following way. We first divide the system into two subsystems (denoted as A and B) and then analyzing the reduced density matrix in one of the subsystems. In our system, the subsystems can be distinguished by the single particle angular momentum $j_z=m+{1\over 2}$: the Subsystem A includes all the positive $j_z$ states, while the Subsystem B includes all the negative $j_z$ states. The total ground state density matrix is given by $\rho=|G\rangle \langle G|$ with $|G\rangle$ denoting the ground state. By the standard procedure, we trace out the Subsystem B to find the reduced density matrix for Subsystem A:
\begin{eqnarray}
\rho^A=\sum_{n_{-j_c},...,n_{-{1\over 2}}}\langle n_{-j_c},n_{-j_c+1},...,n_{-1/2}| \rho | n_{-j_c},n_{-j_c+1},...,n_{-1/2}\rangle,
\end{eqnarray}
where $j_c=m_c+1/2$ denotes a finite-size cutoff of this system. The eigenvalues of the reduced density matrix $\rho_i^A$ give rise to the entanglement spectrum $\xi_i=-\ln \rho_i^A$. For pure Fock state without any correlation, there will be only one non-zero eigenvalue $\rho_i^A=1$ and all the others equal to zero. Therefore, we can observe only one point with $\xi_i\sim 0$ and other points $\xi_i\gg 1$ in the entanglement spectrum for less correlated ground state. While for strongly correlated ground state, the entanglement spectrum has a broad and flat structure. We can further calculate the ground state EE by $EE=-tr \rho^A \ln \rho^A=-\sum_i \rho_i^A \ln \rho_i^A $. We will find $EE\sim 0$ for less correlated ground state while $EE\gg 1$ for strongly correlated ground state.

\subsection{Hartree-Fock equation for trapped spin-orbit coupled Fermi gas}
For weakly correlated states, the mean-field Hartree-Fock (HF) approximation captures the key physics. The HF approximation neglects quantum correlations of the state by assuming:
\begin{eqnarray}
\Psi_{\rm HF}={1\over \sqrt{N!}} \sum_P (-1)^{P} \phi_1(\vec{r}_1)\phi_2(\vec{r}_2)...\phi_N(\vec{r}_N),
\end{eqnarray}
where $P$ denotes all permutations, and $\phi_\alpha$'s are orthonormal single-particle orbitals to be determined. With this assumption, we can obtain the HF Hamiltonian as follows (adopting the trap units):
\begin{eqnarray}
H_{\rm HF} &=&\int d^2 \vec{r} \,\psi ^{\dag }\left[ -\frac{1}{2}\nabla ^{2}+i\tilde{\lambda}%
\left( -\partial _{y}\sigma _{x}+\partial _{x}\sigma _{y}\right) +\frac{1}{2}%
r^{2}+\frac{g }{4}n\!\left( r\right) -\frac{g }{4}\vec{m}\!\left(
r\right) \cdot \vec{\sigma}\right]\, \psi,
\end{eqnarray}
where the constant terms $\left(g/4\right) \int dr [\vec{m}(r)^{2}-n(r) ^{2}]$ has been dropped, and $n({r})$, $\vec{m}({r})$ are respectively the averaged local density and spin vecotr:
\begin{eqnarray}
n\left( r\right) =\sum_{\alpha =1}^{N} |\phi _{\alpha
}\!\left( \vec{r}\right)|^2\,,\;\;\;\; \vec{m}\!\left( r\right) = \sum_{\alpha =1}^{N}\phi _{\alpha }^{\dag } (\vec{r})\, \vec{
\sigma}\,\phi _{\alpha }\!\left( \vec{r}\right) \,.
\label{nm}
\end{eqnarray}
To manipulate the interaction term in a spin rotational invariant way, we have
rewritten the interaction term as: $\frac{g}{8}\int dr\left( n^{2}-\vec{s}%
^{2}\right) $ in the above calculation. The single-particle wave functions $\phi_{\alpha}$ with $\alpha=1,2,...,N$ satisfy the HF equations:
\begin{equation}
\left[ -\frac{1}{2}\nabla ^{2}+i\tilde{\lambda}\left( -\partial _{y}\sigma
_{x}+\partial _{x}\sigma _{y}\right) +\frac{1}{2}r^{2}+\frac{g }{4}%
n\!\left( r\right) -\frac{g }{4}\vec{m}\!\left( r\right) \cdot \vec{\sigma}%
\right]\, \phi _{\alpha }\!\left( \vec{r}\right) =\xi _{\alpha }\,\phi _{\alpha }\!\left(
\vec{r}\right),
\end{equation}
which, together with Eq.~(\ref{nm}), form a closed set and can be solved self-consistently.

In our calculation, $\phi_\alpha$'s are expanded onto the single-particle eigenstates defined in Eq.~(\ref{single}): $\phi _{\alpha }\!\left( r\right) =\sum_{nm}u_{\alpha ;nm}\,\Phi
_{n,m}\!\left( r\right) $. Note that in our HF calculation, we do not restrict to the LLL. So we have to introduce a cutoff $N_c$ for quantum number $n$, in addition to the cutoff for quantum number $m$. Under this expansion, the HF equations take the form:
\begin{equation}
\sum_{n_{2}=1}^{N_{c}}\left( \epsilon _{n_{1}m}\delta _{n_{1}n_{2}}+\frac{%
g }{4}N_{n_{1}n_{2}}^{m}-\frac{g }{4}S_{n_{1}n_{2}}^{m}\right)
u_{\alpha ;n_{2}m}=\xi _{\alpha }u_{\alpha ;n_{1}m},  \label{eigenequation}
\end{equation}
where $N_{n_{1}n_{2}}^{m}=\int d^2 \vec{r}\,\Phi _{n_{1},m}^{\dag }\Phi _{n_{2},m}n\left(
r\right) ,S_{n_{1}n_{2}}^{m}=\int d^2\vec{r}\,\Phi _{n_{1},m}^{\dag } \vec{\sigma}\Phi _{n_{2}m} \cdot \vec{m}\left(r\right)$.
Due to the rotational symmetry, $m$ is a conserved quantum number. The Hartree-Fock wave function $\Psi_{\rm HF}$ would be obtained through iteratively solving the above equations until self consistency is reached. While the ED calculation can only hand a few particle numbers (up to 8 in our calculation), we have done HF calculations up to 200 particle number. From our calculation, we found that the mean-field critical interaction strength at which the non-magnetic state changes to ferromagnetic state roughly scales as $g_{\rm HF} \propto N/\tilde{\lambda}^2$.

\subsection{Time-dependent Hartree-Fock theory}
To study the dynamics, we extend the HF calculation to time-dependent situation.
The time-dependent Hartree-Fock equations take the form:
\begin{equation}
\left( -\frac{1}{2}\nabla ^{2}+i\tilde{\lambda}\left( -\partial _{y}\sigma
_{x}+\partial _{x}\sigma _{y}\right) +V(\vec{r},t) +\frac{\kappa }{4}\left[ n\left( r,t\right) -\vec{m}\left(
r,t\right) \cdot \vec{\sigma}\right] \right) \phi _{\alpha }\left(
\vec{r},t\right) =i\partial _{t}\phi _{\alpha }\left( \vec{r},t\right)\,,
\end{equation}
where we have assumed that the trapping potential $V(\vec{r},t)$ is time-dependent.
The initial wave function at $t=0$ is taken as the ground state wave function under $V(\vec{r},t=0)$. The orthonormality of the single-particle orbitals $\phi_\alpha(\vec{r},t)$ at time $t$ is guaranteed by the unitary time evolution. The local density and spin vector $n$ and $\vec{m}$ are still given by Eqs.~(\ref{nm}) with the explicit time dependence.

\begin{figure}[!h]
\begin{center}
\includegraphics[width=4in]{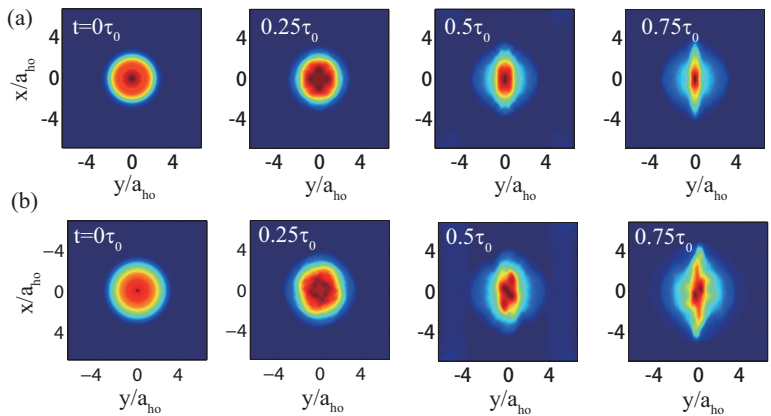}
\end{center}
\caption{(Color online) (Color online) Time evolution of the atomic cloud after a suddent trap deformation. At $t=0$, the trapping frequency along the $y$-axis is suddently changed from $\omega$ to $\sqrt{10} \omega$, while that along the $x$-axis remains at $\omega$. The upper (lower) panel shows the dynamics of a non-magnetic (ferromagnetic) state. Here $\tau_0=1/\omega$, $N=200$ and $\tilde{\lambda}=20$.}
\label{fig:dynamic}
\end{figure}

In the main text, we presented results for 6 fermions.
In Fig. \ref{fig:dynamic}, we show the similar time evolution for 200 fermions. The key physics remains the same: Under a sudden deformation of the trapping potential, the non-magnetic state exhibits quandrupole oscillation, while the chiral ferromagnetic state exhibits an additional rotation analogous to the scissors mode.

\subsection{Mapping to Dresselhaus SO coupling}
In our calculation, we have taken the SO coupling to be of Rashba form: $\mathcal{V}_{\rm soc}=\lambda (p_y\sigma_x-p_x\sigma_y)$. The results can be easily generalized if the SO coupling is of Dresselhaus form: $\mathcal{V}_{\rm SO}^D=\lambda(p_y\sigma_x+p_x\sigma_y)$.
The system with Dresselhaus SO coupling can be mapped to a system with Rashba SO coupling through a unitary transformation in spin space:
$U=i\sigma_x$, under which the Pauli matrices are transformed as \[\sigma_x\rightarrow \sigma_x\,, \;\sigma_y\rightarrow -\sigma_y\,,\; \sigma_z\rightarrow -\sigma_z\,,\] and the Rashba SO coupling is then transformed to the Dresselhaus form. The $s$-wave interaction is spin $SU(2)$ invariant and will not be changed under the above unitary transformation. So, all the results achieved in our main text hold for Dresselhaus SO coupling case after this spin space transformation. For example, the single-particle wave function is obtained by $i\sigma_x\Phi_{n,m}(\vec{r})$, the single-particle and many body energy spectra are unchanged. The main difference is the change of the ground state spin texture, where the local spin transforms as: \[s_x\rightarrow s_x\,, \;s_y\rightarrow -s_y\,, \; s_z\rightarrow -s_z\,,\] as schematically shown in Fig.~\ref{fig:Dresselhaus}. The density current operator coming from the Dresselhaus SO coupling is also modified to be $\vec{j}_s^{(D)}=\lambda (s_y,s_x,0)$, which keeps the density current in the ground state invariant (Fig.~\ref{fig:Dresselhaus}).

\begin{figure}[!h]
\begin{center}
\includegraphics[width=3in]{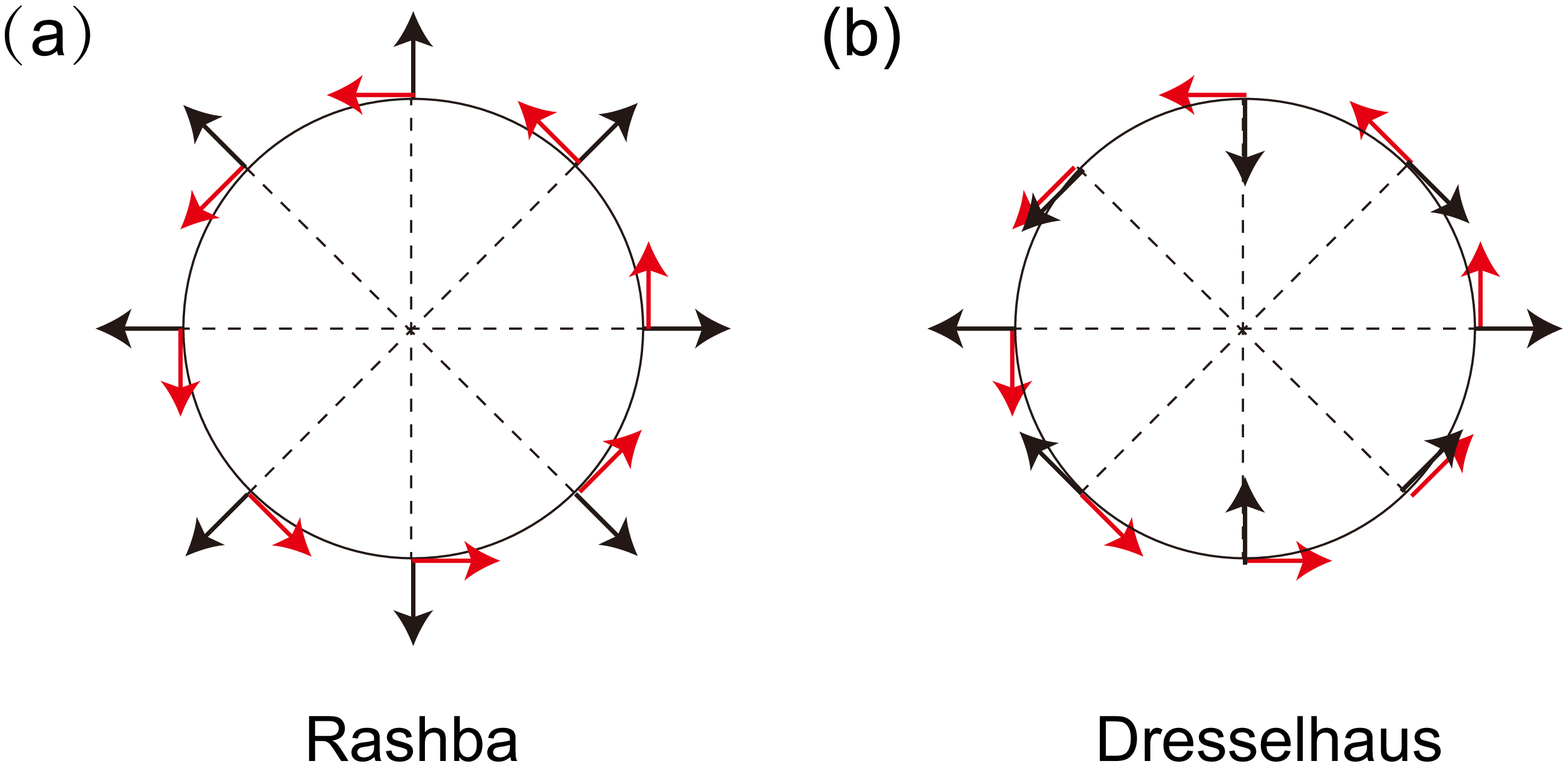}
\end{center}
\caption{(Color online) The schematic plot of the spin polarization (black arrow) and density current (red arrow) for (a) Rashba SO coupled system and (b) Dresselhaus SO coupled system, respectively.}
\label{fig:Dresselhaus}
\end{figure}

\end{widetext}


\begin{thebibliography}{99}

\bibitem{stoner}E. Stoner, Philos. Mag. {\bf 15}, 1018 (1933).

\bibitem{ferro}G.-B. Jo, Y.-R. Lee, J.-H. Choi, C. A. Christensen, T. H. Kim, J. H. Thywissen, D. E. Pritchard, and W. Ketterle, Science {\bf 325}, 1521 (2009).
\bibitem{noferro}C. Sanner, E. J. Su, W. Huang, A. Keshet, J. Gillen, and W. Ketterle, Phys. Rev. Lett. {\bf 108}, 240404 (2012).

\bibitem{gutz}M. C. Gutzwiller, Phys. Rev. Lett. {\bf 10}, 159 (1963); Phys. Rev. {\bf 137}, A1726 (1965).
\bibitem{zhai}H. Zhai, Phys. Rev. A {\bf 80}, 051605(R) (2009).

\bibitem{SOC1} J. Dalibard, F. Gerbier, G. Juzeli\=unas and P. \"{O}hberg, Rev. Mod. Phys. \textbf{83}, 1523 (2011).

\bibitem{SOC11}N. Goldman, G. Juzeli\=unas, P. \"{O}hberg and I. B. Spielman, Rep. Prog. Phys. {\bf 77}, 126401 (2014).

\bibitem{SOC2} Y. J. Lin, K. Jim\'{e}nez-Garc\'{i}a and I. B. Spielman, Nature (London) \textbf{471}, 83 (2011).

\bibitem{SOC3} P. Wang, Z. Q. Yu, Z. Fu, J. Miao, L. Huang, S. Chai, H. Zhai and J. Zhang, Phys. Rev. Lett. \textbf{109}, 095301 (2012).

\bibitem{SOC4} L. W. Cheuk, A. T. Sommer, Z. Hadzibabic, T. Yefsah, W. S. Bakr and M. W. Zwierlein, Phys. Rev. Lett. \textbf{109}, 095302 (2012).


\bibitem{SOC5} J. Zhang, H. Hu, X. J. Liu and H. Pu, in {\em Annual Review of Cold Atoms and Molecules} (World Scientific, Singapore, 2014), edited by K. Madson, K. Bongs, L. D. Carr, H. Zhai, and A. M. Rey.

\bibitem{SOC6} H. Zhai, Rep. Prog. Phys. \textbf{78}, 026001 (2015).


\bibitem{wu}C. Wu, I. Mondragon-Shem, and X.-F. Zhou, Chin. Phys. Lett. {\bf 28}, 097102 (2011).

\bibitem{HanPu1} H. Hu, B. Ramachandhran, H. Pu, and X. -J. Liu, Phys. Rev. Lett. \textbf{108}, 010402 (2012).

\bibitem{HanPu2} B. Ranachandhran, H. Hu, and H. Pu, Phys. Rev. A \textbf{87}, 033627 (2013).



\bibitem{supp}See SupplementalMaterial for details of the ED calculation, the calculation of entanglement entropy, the time-independent and the time-dependent HF calculation, and the discussion about similarities and differences between Rashba and Dresselhaus SO coupling.

%

\bibitem{conduit} G. J. Conduit, Phys. Rev. A {\bf 82}, 043604 (2010).
\bibitem{toigo}A. Ambrosetti, G. Lombardi, L. Salasnich, P. L. Silverstrelli, and F. Toigo, Phys. Rev. A {\bf 90}, 043614 (2014).

\bibitem{pilati}S. Pilati, I. Zintchenko, and M. Troyer, Phys. Rev. Lett. {\bf 112}, 015301 (2014).

\bibitem{sci1}C. Lobo, A. Sinatra, and Y. Castin, Phys. Rev. Lett. {\bf 92}, 020403 (2004); C. Lobo, and Y. Castin, Phys. Rev. A {\bf 72}, 043606 (2005).

\bibitem{shanxi}L. Huang, Z. Meng, P. Wang, P. Peng, S.-L. Zhang, L. Chen, D. Li, Q. Zhou, and J. Zhang, arXiv:1506.02861.






\end{thebibliography}
\end{document}